\documentstyle[prd,aps,floats]{revtex}

\begin{document}

\draft
\input epsf

\title{
Polar Perturbations of Self-gravitating 
Supermassive Global Monopoles
} 

\author{
Hiroshi Watabe\thanks{watabe@gravity.phys.waseda.ac.jp} 
and
Takashi Torii\thanks{torii@gravity.phys.waseda.ac.jp}
}

\address{
Advanced Research Institute for Science and Engineering,
Waseda University, Shinjuku-ku, Tokyo 169-8555, Japan
}

\date{\today}

\maketitle

\begin{abstract}
Spontaneous global symmetry breaking of $O(3)$ scalar field 
gives rise to point-like topological defects, global monopoles.
By taking into account self-gravity,
the qualitative feature of the global monopole
solutions depends on the 
vacuum expectation 
value $v$ of the scalar field. 
When $v<\sqrt{1/8\pi}$, there are global monopole solutions
which have a deficit solid angle defined at infinity.
When $\sqrt{1/8\pi}\leq v <\sqrt{3/8\pi}$, 
there are global monopole solutions with the cosmological horizon,
which we call the supermassive global monopole.
When $v\geq\sqrt{3/8\pi}$,  there is no nontrivial solution.
It was shown that all of these solutions are stable against the
spherical perturbations. 
In addition to the global monopole solutions,
the de Sitter solutions exist for any value of $v$. They are
stable against  the spherical perturbations when $v\leq\sqrt{3/8\pi}$,
while unstable for $v>\sqrt{3/8\pi}$.
We study polar perturbations of these solutions and
find that all self-gravitating global monopoles are stable even
against polar perturbations, independently of the
existence of the cosmological horizon, while
the de Sitter solutions are always unstable. 

\end{abstract}



\setcounter{footnote}{1}
\renewcommand{\thefootnote}{\fnsymbol{footnote}}

\section{Introduction}
\label{Intro}

Phase transitions in the early universe are caused by the symmetry breaking
leading to a manifold of degenerate vacua with nontrivial topology and
giving rise to topological 
defects.  The defects are classified by the topology of the vacua such into
domain walls, cosmic strings and monopoles.
If the gauge field is involved in the spontaneous symmetry breaking,
the defects are gauged. On the other hand, when the symmetry
is global, the emerging defects are called global defects. 

In this paper, we shed light on  global monopoles. 
Although energy of  the gauge 
monopoles is finite,  the global monopoles have divergent energy
because of the long tail of the field.
This divergence has to be removed by cutting off at a certain distance. 
This procedure is not necessarily artificial, because another defect 
which may exist near the original one cancels the divergence. 
This secondary defect is not only the monopole, but also can be 
a domain wall or a cosmic string.

Global monopoles have been an interesting subject in cosmology.  They 
were thought as of the seeds of structure formation or inflation.   
By taking into account the self-gravity of the global monopoles, 
it can give rise to a deficit solid
angle\cite{Barriola}, which would affect cosmological data. It, 
however, may be counteracted when the universe has a cosmological 
constant. 

Vilenkin and Linde independently pointed out that topological defects
can cause inflation \cite{Vilenkin,Linde}. When the vacuum 
expectation value (VEV) is larger than a certain critical value,
the scalar field stays on the top of the potential 
after global symmetry breaking. 
In this case, the space would expand exponentially with time. 
This inflation model is called topological inflation.
Topological inflation does not suffer from the initial 
value problem, unlike the new or hybrid inflation. 
Sakai et al. \cite{Sakai}
found the critical vacuum expectation value (VEV) is $0.33M_P$ numerically. 

Recently new types of the self-gravitating global monopole solutions were
discovered numerically\cite{Liebling}.  As the VEV of the
$O(3)$ scalar field $v$ increases, the deficit solid angle also 
gets large and it becomes $4\pi$ when $v =v_{cri}:=\sqrt{1/8\pi}$.
Beyond this critical value there is no ordinary monopole solution, but
there appears a new type of solution in the parameter range 
$v_{cri} < v < v_{max} :=\sqrt{3/8\pi}$.
This has a cosmological horizon at $r=R_c$ and gives a natural cutoff scale.
The appearance of the new solution is similar to the supermassive
string solution\cite{string}, which has a deficit  angle larger than $2\pi$.
In this sense, we will call this solution the supermassive global monopole in 
contrast to the ordinary global monopole.

From gravitational theoretical interest, the black hole counterpart
of the global monopole is also discussed. Maison investigated  the scalar
hair of the black holes in the same system and showed their
existence domains as a function of the radius of the event horizon\cite{MaisonBH}.
Nucamendi and Sudarsky discussed the definition of the Arnowitt-Deser-Misner (ADM) mass
in spacetime with deficit solid angle and found that it becomes
negative for small  black holes\cite{Nucamendi}.

One of the most important issues of these kinds of  isolated objects
is  stability.  In the ordinary monopole case without gravity,
if Derrick's no-go theorem\cite{Derrick} could be applied,
they would be unstable towards radial rescaling
of the field configuration. This is, however, not the case due to
the diverging energy of the solutions. It was demonstrated that the
ordinary monopole solutions are stable against spherical perturbations.
As for the non-spherical perturbations, there was some debate. 
Goldhaber\cite{Goldhaber} investigated the polar perturbations and found that 
the energy functional is the same form as that of the sine-Gordon equation
under some conditions. Hence there would be a zero mode leading
a north-pointing teardrop shape for the monopole mass density and
the knot would be untied.
Rhie and Bennett\cite{Rhie}, however, pointed out that such instability
is just the artificial fixing of the monopole core. If the monopole is
free to move, which is the natural situation of the isolated system,
the monopole cannot become a teardrop shape and is stable.
This was actually confirmed by numerical 
simulations\cite{Perivio,Achucarro}.
Bennett and Rhie\cite{Bennett} also pointed out 
by numerical calculation with 2D code
that a monopole and antimonopole pair would collapse to a string 
through the unwinding process when these cores are artificiality fixed.
If these cores of the pair are free to move, 
such an unwinding process does not occur.

It is expected that the self-gravitating global monopoles with
$v <v_{cri}$ (i.e., without cosmological horizon) have
the same stability properties as a non-gravitating one.
Stability may change, however,  for the supermassive monopole
solutions ($v >v_{cri}$), because the domain of 
communication, that is, the boundary condition, is different
due to the cosmological horizon. 
Maison and Liebling investigated the spherical perturbations of the
supermassive 
monopole solution\cite{Maison}. They make use of  de Sitter
solutions, which are trivial solutions such that the scalar field
stays at the top of the potential barrier and exists for any value of 
the VEV of the scalar field. 
By their analysis the stability
change of the de Sitter solutions occurs at $v =v_{max}$,
beyond which the solutions are stable, while unstable below that.
And the supermassive monopole solutions emerge just at this value
if the VEV decreases from a larger value. This kind of behavior
can be seen in a variety of systems in nature and explained
by using catastrophe theory.  The supermassive monopole solutions inherit
the stability from the de Sitter solution with $v > v_{max}$.
As a result, they are stable against the spherical perturbations
even if they have a cosmological horizon.

Then, are the supermassive monopole solutions really stable? We have to 
examine this question carefully. This is because the polar perturbation pushes
the scalar field configuration of the de Sitter solution 
to one direction in the internal space 
from the top of the potential barrier. In the anti-de Sitter background,
such a solution can be stable even in the tachyonic situation
if the effective mass of the scalar field satisfies the 
Breitenlohner-Freedman bound\cite{Breitenlohner}.
In the de Sitter case, however, it is easy to imagine the scalar field
rolls down to its VEV, i.e., the solution is unstable.
Hence, the supermassive monopole solutions may be unstable against the
polar perturbations. To settle this issue is the main purpose of
this paper.

This paper is organized as follows. 
In Sec.~\ref{sec:sphe} we review the 
static solutions and their spherical stability. 
In Sec.~\ref{sec:ds}, we show the instability of the de Sitter solution 
against the polar perturbations.
In Sec.~\ref{sec:polar}, we formulate the polar perturbations of the $O(3)$ 
scalar field.
In Sec.~\ref{sec:stability}, we show the stability 
of the self-gravitating global monopoles. 
Throughout this  paper, we use the units $\hbar=c=G=1$.


\section{Static solutions}
\label{sec:sphe}
In this section, we briefly review the self-gravitating
global monopole solutions\cite{Barriola,Maison}.
The theory of a scalar field with spontaneously broken internal
$O(3)$ symmetry, minimally coupled to gravity, is described
by the action
\begin{equation}
S=\int d^4x \sqrt{-g}\left[\frac{R}{16\pi}
-\frac{1}{2} \partial_{\mu}\Phi^{a} \partial^{\mu} 
\Phi^{a}-\frac{\lambda}{4}(\Phi^a \Phi^a-v^2)^2 \right],
\end{equation}
where $R$ is the Ricci scalar of the spacetime and $\Phi^{a}$ $(a=1,2,3)$
is the triplet scalar field.
$\lambda$ and $v$ are the self-coupling constant and
the VEV of the scalar field, respectively.
The energy momentum tensor is 
\begin{equation}
T_{\mu \nu}=\partial_{\mu}\Phi^{a} \partial_{\nu}\Phi^{a}-g_{\mu \nu}
\left[\frac{1}{2}\partial_{\rho}\Phi^{a}\partial^{\rho}\Phi^{a}
+\frac{\lambda}{4}(\Phi^2-v^2)^2 \right].
\end{equation}
For the static solution with unit winding number, we adopt the
so-called hedgehog ansatz
\begin{equation}
\Phi^{a}=h(r)\frac{x^a}{r},
\end{equation}
where $x^a$ are the Cartesian coordinates.

We shall consider the static spherically symmetric spacetime and adopt 
a Schwarzschild type metric 
\begin{equation}
ds^2=-f(r)e^{-2\delta(r)}dt^2 +\frac{1}{f(r)}dr^2 
+r^2 (d \theta^2 +\sin^2\theta d \phi^2),
\end{equation}
where 
\begin{eqnarray}
f(r)=1-\frac{2m(r)}{r}.
\end{eqnarray}

Under these Ans\"atze, we get these field equations,
\begin{equation}
m'
=4 \pi r^2\left[\frac{1}{2}fh'{}^2
+\frac{h^2}{r^2}+\frac{\lambda}{4}(h^2-v^2)^2\right],
\end{equation}
\begin{equation}
\delta'=-4\pi r h'{}^2,
\end{equation}
\begin{equation}
\frac{1}{r^2e^{-\delta}}\left[r^2 e^{-\delta}fh'\right]'-\frac{2h}{r^2}
=\lambda (h^2-v^2)h,
\end{equation}
where a prime denotes a derivative with respect to the radial coordinate.

These equations are integrated with suitable boundary conditions.
At the center the spacetime should be regular.
By expanding these equations, we find 
$h'(0)$ can be regarded as a free parameter, 
which is determined by the other boundary 
condition at $r \to \infty$ (for the ordinary global monopole case) or 
$r = R_c$  (for the supermassive global monopole case).
For the ordinary global monopole solution the spacetime 
approaches asymptotically
flat spacetime (which implies that the curvature vanishes) with  deficit solid angle $\alpha$
\begin{eqnarray}
f &\to& 1-\alpha -\frac{2M}{r}+O\left(\frac{1}{r^2}\right),
\\
\delta &\to& O\left(\frac{1}{r^4}\right),
\\
h &\to& v +O\left(\frac{1}{r^2}\right).
\end{eqnarray}
On the other hand, we impose the
existence of the regular cosmological horizon at $r=R_c$ for the 
supermassive global monopole.

There is a trivial de Sitter  solution 
$h(r)\equiv 0$, $f(r)= 1-r^2/R_c^2$, and $\delta(r)\equiv 0$.
This solution has a cosmological horizon at 
$r=R_c=\sqrt{3/\Lambda_{eff}}$,
where the effective cosmological constant is 
$\Lambda_{eff}:=2\pi\lambda v^4$.
These solutions exist for any value of $v$.

\begin{figure}[t!]
\centering
\hspace*{-7mm}
\leavevmode
\epsfysize=5cm 
\epsfbox{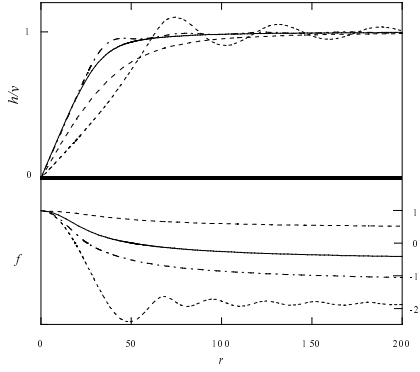}
\\[2mm]
\caption[fig-1]{\label{fig:static}
The field configurations (above) and the metric function f(r) (below) of the
global monopole solutions for $v=0.15$ (dashed line), $0.25$ (solid line),
$0.30$ (dot-dashed line) and $0.34$ (dotted line). 
The solutions except for $v=0.15$
 have a cosmological horizon. The solutions for $v>\sqrt{2/8\pi} \approx 0.28$ 
 show oscillating behavior.
}
\end{figure}

The ordinary global monopole solutions exist for 
$0<v<v_{cri}$. The configuration of the scalar
field is shown in Fig.~\ref{fig:static} ($v =0.15$). 
We set $\lambda = 0.1$
(which is adopted in Ref.~\cite{Liebling}) without loss of generality
throughout this paper
since $\lambda $ can be scaled out by introducing new variables
$\bar{r}:=\lambda^{1/2}v r$, $\bar{m}:=\lambda^{1/2} v m$ and 
$\bar{\Phi}^a:=\Phi^a/v$.
The deficit solid angle $\alpha$ becomes large as $\alpha=4\pi(8\pi v^2)$,
and $\alpha=4\pi$ for $v=v_{cri}$, which
implies the disappearance of the asymptotic region.

\begin{figure}[t!]
\centering
\hspace*{-7mm}
\epsfysize=5cm 
\epsfbox{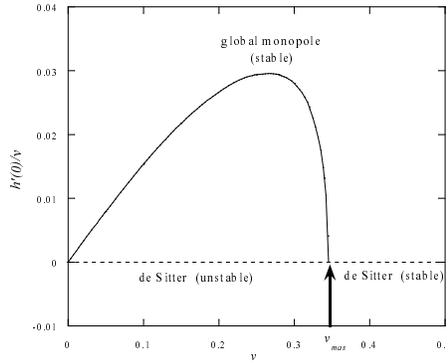}
\\[2mm]
\caption[fig-1]{\label{fig:dh0}
The diagram of  $v$ versus $h'$ at $r=0$.
The solid and the dashed line represent the monopole and the de Sitter
solutions, respectively. 
The stability here means against the spherical perturbations. 
This figure says that the monopole solutions and 
de Sitter solution with $v<v_{max}$ are  stable against the spherical 
perturbations. 
$v_{max}$ is the bifurcation point which connects the stable 
de Sitter, the global monopole, and the unstable de Sitter solutions.
}
\end{figure}

Beyond the critical value the supermassive global monopole
solutions appear for $v_{cri}<v<v_{max}$~\cite{Liebling}.
This has a cosmological horizon. If $\sqrt{2/8\pi}<v<v_{max}$,
the scalar field shows the oscillating behavior approaching its VEV
beyond the  cosmological horizon (See Fig.~\ref{fig:static}). 
Asymptotically it becomes the de Sitter spacetime~\cite{Maison}.
At $v=v_{max}$ the solution coincides continuously
(at least in the domain of the communications) 
with the de Sitter solution. This can be seen in the behavior
of the parameter $h'(0)$ as shown in Fig.~\ref{fig:dh0}.

Maison and Liebling investigated the stability of the de Sitter solutions
against spherical (both in spacetime and in internal space) perturbations
and found that stability changes at $v=v_{max}$.
By this result they expected that the stable property of the de Sitter
solution with $v>v_{max}$ is transferred to the supermassive
global monopole solutions. This kind of study was performed by 
using catastrophe theory and applied for the black hole 
spacetimes\cite{torii,tachi}.
However, are the supermassive global monopoles really stable even
for the non-spherical perturbations? The O(3) field 
which constructs the de Sitter solution is
always on the top of the Mexican hat potential, which seems unstable.  
It is easily imagined that the scalar field  rolls down
to their VEV by just pushing it in one direction in the internal space.
We investigate this problem in the next section.

\section{Instability of de Sitter solutions} 
\label{sec:ds}

Now, we investigate  time dependent 
perturbations of the de Sitter solutions which are non-spherical
in the internal space. 
We perturb  one  component of the $O(3)$ scalar field
\begin{equation}
\Phi^1(t, r)\equiv \Phi^2(t, r)\equiv0, \;\;\; \Phi^3(t, r)=e^{i\sigma t}\zeta(r).
\end{equation}
Note that the metric functions are not affected by this perturbation
since the back reaction is second order, and  we can consider the de
Sitter spacetime as a background.
If the perturbation equation allows a solution with 
imaginary $\sigma$, the mode function  
evolves exponentially with time. That indicates instability of the 
de Sitter solution.

The perturbation equation is written as,
\begin{equation}
\frac{1}{r^2}\left[r^2f\zeta'\right]\!'
+\lambda v^2 \zeta
=-f^{-1}\sigma^2\zeta.
\end{equation}
Adopting the tortoise coordinate $r_{*}$,
\begin{eqnarray}
\frac{dr}{dr_{*}}=f,
\end{eqnarray}
and a new variable $\bar{\zeta}=r\zeta$,
we can rewrite it  in the Schr\"{o}dinger equation,
\begin{equation}
-\frac{d^2 \bar{\zeta}}{dr_{*}^2}+U(r)\bar{\zeta}=\sigma^2\bar{\zeta}.
\end{equation}
$U(r)$ is the potential function of the linear equation
\begin{equation}
U(r)=-f \lambda v^2 \left(1+\frac{4}{3}\pi v^2\right).
\end{equation}
The potential function never becomes positive inside the cosmological horizon. 
This is the exactly the same form as the Eqs.~(37)-(39) in Ref.~\cite{torii2}
in the de Sitter case. In that paper it was shown that these equations 
always have at lease one negative eigenmode for any value 
of $\lambda$ and $v$. 
Hence, it is concluded that all the de Sitter solutions which we consider 
here are unstable.

\section{Polar perturbation of the scalar field}
\label{sec:polar}

Since  the de Sitter solutions have  polar instabilities,
the self-gravitating global monopoles  may inherit them.
Hence we study the polar perturbation of the global monopoles in general relativity
carefully in this section and the following section. 

The polar deformation of the $O(3)$ scalar field was proposed 
by Goldhaber\cite{Goldhaber}. He 
introduced a new coordinate $y:=\ln \tan(\theta/2)$ to discuss the invariance 
of the energy under the deformation. Ach\'ucarro and Urrestilla improved his 
notation\cite{Achucarro},
\begin{eqnarray}
\Phi^{1}&=&H(t,r,\theta)\sin \bar{\theta}(t,r,\theta) \cos \phi,  
\nonumber \\
\Phi^{2}&=&H(t,r,\theta)\sin \bar{\theta}(t,r,\theta) \sin \phi, \\
\Phi^{3}&=&H(t,r,\theta)\cos \bar{\theta} (t,r,\theta), \nonumber
\end{eqnarray}
and
\begin{eqnarray}
\tan(\bar{\theta}/2)=e^{y+\xi(t,r,\theta)}.
\end{eqnarray}
$\bar{\theta}$ is the polar component of $\Phi^a$ and 
$H(t,r,\theta)=h(r)+\delta h(t,r,\theta)$.
When $\xi=0$, i.e., $\bar{\theta}=\theta$ and 
$\delta h = \delta h(t, r)$ the global 
monopole is spherically symmetric, while it become a `string'
when $\xi \rightarrow \infty$.
The energy of the static global monopole is expressed with new coordinate 
$y$,
\begin{eqnarray}
E&=&\int drdyd\phi (\rho_1+\rho_2),
\end{eqnarray}
where
\begin{eqnarray}
\rho_1&=&\frac{H^2}{2}\left[\sin^2\bar{\theta}
+\left(\frac{\partial \bar{\theta}}{\partial y}\right)^2 
+\frac{r^2}{\cosh^2 y}
\left(\frac{\partial \bar{\theta}}{\partial r} \right)^2\right],
\\
\rho_2&=&\frac{1}{2}\left(\frac{\partial H}{\partial y}\right)^2
+\frac{r^2}{2\cosh^2 y}
\left[
\left(\frac{\partial H}{\partial r}\right)^2
+\frac{1}{2}(H^2-v^2)^2\right].
\nonumber \\
\end{eqnarray}
In the far region from the monopole core, the terms  
$\partial H/\partial r$ and $\partial H/\partial y$ can be disregarded. 
Goldhaber pointed out that the 
first two terms in $\rho_1$ are in the same form as the 
energy of a sine-Gordon soliton, 
which implies that the energy is invariant under translation of 
coordinate $y$ (or $\xi$). 
Thus, he concluded  that the global monopoles have instability 
if there is a deviation in which 
$H=v$ and $\partial \bar{\theta}/\partial r=0$ are held. 
Rhie and Bennett proved, however, that such a collapse does not occur, if the
monopole core is free to move\cite{Rhie}. 

We assume $\xi$ and $\delta h$ are small. Therefore we take
\begin{eqnarray}
\bar{\theta}&=&\theta+\xi \sin \theta.
\end{eqnarray}

Now we have two perturbative functions $\xi$ and $\delta h$ for
the $O(3)$ scalar field.
The simplest polar perturbation assumes that $\delta h=0$ and $\xi$ is 
independent of $\theta$. However, this is inconsistent. 
In a non-relativistic case, such perturbation must vanish 
through a perturbation equation. 
If $\xi$ has $\theta$ dependence, we
will get
\begin{equation}
\partial_{\theta}\xi+2 \xi \cot \theta =0.
\end{equation}
The solution of this equation is $\xi \propto (\sin\theta)^{-2} $,
which is unphysical because  $\xi$ violates the perturbative
approximation around the axis.

The same feature appears in the case that only $\delta h$ 
is considered. 
Consequently, we can conclude that perturbation only with 
either $\xi$ or 
$\delta h$  does not occur when the self-gravity is set to zero. 
We will find that this is 
true also in the self-gravitating case.

Now, we assume $\xi$ is independent of $\theta$. 
Since we will examine the $l=1$ Legendre type perturbation below,
we take $\delta h = \eta(r)\cos \theta e^{i\sigma t}$. 
The perturbed scalar field is
\begin{eqnarray}
\Phi^{1}&=&(h+\eta \cos \theta e^{i \sigma t})\sin 
\theta \cos \phi+h\xi e^{i \sigma t} \sin \theta \cos \theta 
\cos \phi,  
\nonumber \\
\Phi^{2}&=&(h+\eta \cos \theta e^{i \sigma t} )
\sin \theta \sin \phi+h\xi e^{i \sigma t} \sin \theta 
\cos \theta \sin \phi, 
\nonumber
\\
\Phi^{3}&=&(h+\eta \cos \theta e^{i \sigma t} )
\cos \theta -h\xi e^{i \sigma t} \sin^2\theta \nonumber.
\\
\label{eqn:ptb}
\end{eqnarray}

Let us consider the combination
\begin{eqnarray}
p(r)=h\xi+\eta, \;\;\; q(r)=h\xi.
\end{eqnarray}
in Eq.~(\ref{eqn:ptb}).
Putting $p(r)\equiv 0$, we recover the polar perturbation of the
de Sitter solution
\begin{eqnarray}
\Phi^1\equiv \Phi^2\equiv0, \;\;\;
\Phi^3=-q(r)e^{i \sigma t}.
\end{eqnarray}
which is discussed in Sec.~\ref{sec:ds}.

\begin{figure}[t!]
\centering
\hspace*{-7mm}
\leavevmode
\epsfysize=5cm 
\epsfbox{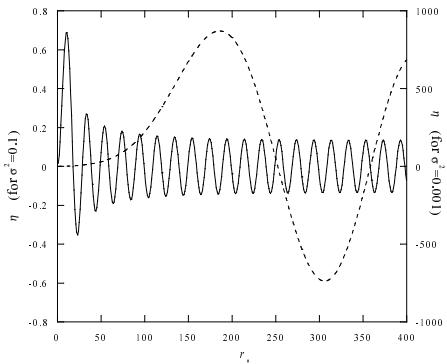}
\\[2mm]
\caption[fig-1]{\label{fig:v025a}
The eigenmodes of $\eta$ and $h \xi/v$ 
for $\sigma=0.1$ (solid line) and $0.001$ (dashed line) 
when $v=0.25$.
We can see the oscillating behavior. These are continuous modes.
}
\end{figure}

\section{polar perturbation of the global monopole}
\label{sec:stability}

Ach\'ucarro and Urrestilla  studied stability of the global monopoles 
against polar 
perturbation with all orders\cite{Achucarro} and 
found that the global monopoles are stable. 
Here we extend their analysis to the self-gravitating cases.

For the global monopole solutions, the matter field
is non-zero. So the first order perturbations of the matter field
couple to 0-th order
to create the metric perturbations, which can be 
described as
\begin{eqnarray}
ds^2&=&-fe^{-2\delta}e^{\delta \nu}dt^2
+\frac{1}{f}e^{\delta \mu_{2}} dr^2
+r^2e^{\delta \mu_{3}}d \theta^2
+r^2 \sin^2 \theta e^{\delta \psi}d\phi^2.
\end{eqnarray} 
The valuables in the metric can be separated \cite{Friedman},
\begin{eqnarray}
\delta \nu &=&\sum N_{l}(t,r)P_{l}(\cos \theta), \\
\delta \mu_{2}&=&\sum L_{l}(t,r) P_{l}(\cos \theta), \\
\delta \mu_{3}&=&\sum [T_{l}(t,r)P_{l}(\cos \theta) 
+S_{l}(t,r) P_{l,\theta,\theta}(\cos \theta)], \\
\delta \psi&=&\sum [T_{l}(t,r) P_{l}(\cos \theta) 
+S_{l}(t,r)P_{l,\theta}(\cos \theta) \cot \theta],
\end{eqnarray}
where $P_{l}$ is a Legendre polynomial.
Now we  consider the simplest case $l=1$ and drop the suffix $l$. 
For the higher order Legendre polynomials,
the eigenvalue $\sigma^2$ is expected to be larger than that for $l=1$.
Thus, the perturbed 
metric can be written in the form,
\begin{eqnarray}
ds^2&=&-\left[f+B(r)e^{i \sigma t} \cos \theta\right]
e^{-2\delta}dt^2 
+\frac{1}{f}\left[1+L(r)e^{i \sigma t} \cos \theta\right]dr^2 
+r^2\left[1+T(r)e^{i \sigma t}\cos \theta\right]
(d \theta^2+\sin^2 \theta d \phi^2).
\label{eqn:met-ptb}
\end{eqnarray}
Here, we introduced a new variable $B(r):=N(r)f(r)$ for convenience,
re-defined $T := T-S$ and assumed harmonic time dependence.

The energy momentum tensor is, 
\begin{eqnarray} 
{T_{t}}^t&=&\stackrel{\mbox{\tiny $(0)$}}{T}\!^{t}_{t}
+\Biggl[-f h'\eta'+\frac{f h^{'2}}{2}L
-\frac{2h}{r^2}(h\xi+\eta) 
+\frac{h^2}{r^2}T-\lambda h(h^2-v^2)\eta
\Biggr]\cos \theta e^{i\sigma t},
\\
{T_{r}}^r&=&\stackrel{\mbox{\tiny $(0)$}}{T}\!^r_{r}
+\Biggl[f h'\eta'-\frac{f h^{'2}}{2}L
-\frac{2h}{r^2}(h\xi+\eta)
+\frac{h^2}{r^2}T
-\lambda h(h^2-v^2)\eta\Biggr]\cos \theta e^{i \sigma t}, 
\\ 
\nonumber\\
{{T}_{t}}^{r}&=&
i \sigma  fh'\eta \cos \theta e^{i \sigma t}, 
\\
{{T}_{t}}^{\theta}&=&
i \sigma \frac{h^2}{r^2} \xi \sin \theta e^{i \sigma t},
\end{eqnarray}
where $\stackrel{\mbox{\tiny $(0)$}}{T}\!_{t}^{t}$ and
$\stackrel{\mbox{\tiny $(0)$}}{T}\!_{r}^{r}$ are non-perturbed
components.
We displayed only the first order perturbation. Thus, we can get 
the perturbation equations
\begin{equation}
L+T=16 \pi h^2 \xi,
\label{eqn:LT}
\end{equation}
\begin{equation}
f(-rT'+L-rT\delta' -T)+\frac{1}{2}f'rT=8 \pi r f 
h'\eta,
\label{eqn:T'}
\end{equation}
\begin{eqnarray}
&&fr^2T''-(frL)'+\frac{1}{2}f'T'r^2+3frT'-L
=4\pi\left[-2fr^2h'\eta'+fr^2h^{\prime 2}L+2h^2T
-4h(h\xi+\eta) \right.
 \left.
-2r^2\lambda h(h^2-v^2)\eta\right],
\label{eqn:T2}
\end{eqnarray}
\begin{eqnarray}
&& fr^2T''-frL'-L+2frT'-2fr \delta' L+r^2f\delta' T'
-\frac{r^2 \sigma^2e^{2\delta}}{f}T-B'r+\frac{B}{f}+\frac{Brf'}{f}
=8\pi r^2 f h'\left( h' L+2\eta' \right),
\label{eqn:N}
\end{eqnarray}
from the Einstein equation, and
\begin{eqnarray}
&& \frac{\sigma^2h e^{2\delta}}{f}\xi+f(h\xi)''
+\left(\frac{2}{r}f
+f'-\delta'f\right)(h\xi)'
-\frac{2\eta+2h\xi}{r^2}+\frac{hL}{2r^2}
-\frac{1}{2}\frac{hB}{fr^2}=\lambda h(h^2-v^2)\xi,
\label{eqn:phi12}
\end{eqnarray}
\begin{eqnarray}
&& \frac{\sigma^2e^{2\delta}}{f}\eta+f\eta''
+\left(\frac{2}{r}f
+f'-\delta'f\right) \eta'
+fh'T'
-\left[\frac{2h}{r^2}
+\lambda h(h^2-v^2) \right]L
-\frac{4h\xi+4\eta-2hT}{r^2}
\nonumber \\
&& \;\;\;\;
-\frac{fh'L'}{2}+\frac{1}{2}h'B'-\frac{h'f'B}{2f}
=\lambda(3h^2-v^2)\eta,
\label{eqn:phi3}
\end{eqnarray}
from the  equation of the scalar field.

The boundary conditions at $r=0$ are obtained by imposing
regularity. By Taylor expansion around $r=0$, we find
\begin{eqnarray}
\xi&=&\xi_{1}r+O\left(r^3\right),
\\
\eta&=&\frac{h_1}{4}(8\xi_1-B_1)r^2+O\left(r^4\right),
\\
B&=&B_1 r- \frac16\left(m_3+\frac{24}{5}\pi h_1^2\right)B_1r^3
+O\left(r^5\right),
\\
T&=& \frac{2}{5}\pi h_1^2B_1r^3+O\left(r^5\right),
\\
L&=& \left(16\pi \xi_1-\frac{2}{5}\pi B_1\right)h_1^2r^3
+O\left(r^5\right),
\end{eqnarray}
where 
$F_{n}$ represents the n-th order derivative coefficient
of $F$ at $r=0$. We assumed all values at $r=0$ are $0$,
 for the elimination of trivial translations 
mode in which $\eta=-h'\cos \theta $ and $ h\xi=h/r$ \cite{Achucarro}.
This assumption is not, however, an artificial fixing of the core, 
but only coordinate transformation.
The values of $h_1$ and $m_3$ are given by the static solutions
in Sec.~\ref{sec:sphe}.
$\xi_1$ and $B_1$ cannot be determined by this regularity condition. 
Among these, $\xi_1$ is arbitrary because of the freedom of the constant
multiplication in the linear theory. Hence there are two parameters
$B_1$ and $\sigma^2$ in this system, which should be
adjusted to obtain the regular normalizable eigenmodes.
If there is no solution, we cannot 
find an appropriate set of $B_1$ and $\sigma^2$.

By these boundary conditions we solve Eqs. 
(\ref{eqn:LT})-(\ref{eqn:phi3}) numerically. 
Fig.~\ref{fig:v025a}
is the typical solution of the supermassive case.
We can find the oscillating behavior for the positive eigenvalue
$\sigma^2$. These are the continuum modes.

\begin{figure}[t!]
\centering
\hspace*{-7mm}
\leavevmode
\epsfysize=5cm 
\epsfbox{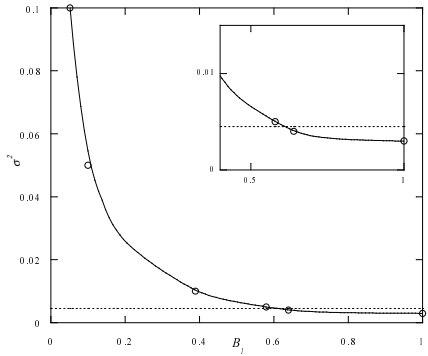}
\\[2mm]
\caption[fig-1]{\label{fig:B1_vs_sig_v015}
The diagram of the shooting parameter $B_{1}$ versus 
$\sigma^2$ when $v=0.15$.  The background global monopole
solution does not have a cosmological  horizon.
We can see $\sigma^2$ converges to positive constant for 
large $B_1$.
The horizontal line is minimum eigenvalue 
$\sigma_{min}:=2\lambda v^2$ in the non-self-gravitating case. 
The asymptotic value of $\sigma^2$ for the self-gravitating case is smaller 
than that of the non-self-gravitating case.
}
\label{fig:B1_vs_sig_v015
}
\end{figure}

If we neglect self-gravity, the perturbation equations 
become quite simple. In this case it is easy to observe that
the potential function is nonzero constant asymptotically
to infinity. Hence there is a minimum value of the eigenvalue
$\sigma_{min}^2=2\lambda v^2>0$ 
for the continuous modes. When 
self-gravity is taken into account, the situation does not change
seriously for the ordinary global monopole solution. 
The existence of the minimum eigenvalue is seen in 
Fig.~\ref{fig:B1_vs_sig_v015}.
Although $\sigma_{min}\!^2$ decreases continuously for large $B_1$,
it converges to a positive constant which is smaller than that of
 the non-self-gravitating case.

\begin{figure}[t!]
\centering
\hspace*{-7mm}
\leavevmode
\epsfysize=5cm 
\epsfbox{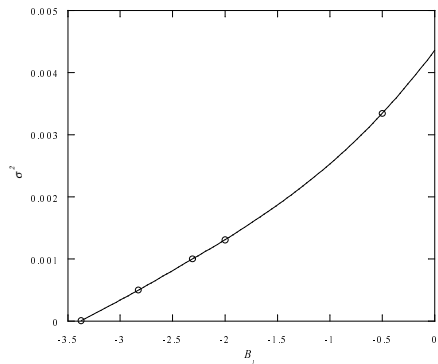}
\\[2mm]
\caption[fig-1]{\label{fig:B1_vs_sig_v025}
Diagram of the shooting parameter $B_{1}$ versus 
$\sigma^2$ when $v=0.25$,
which corresponds to
the case of the  supermassive global monopoles. 
We can see that 
$\sigma^2$ is continuously decreasing to $0$, but 
there is no zero mode as seen in Fig.~\ref{fig:sig0}.
}
\end{figure}

For the supermassive global monopole, however, the form of the
potential function is qualitatively different due to the 
existence of the cosmological horizon.
It vanishes at the horizon without bottom up, and hence
the continuous modes   exist for the infinitesimally
small eigenvalue as seen in Fig.~\ref{fig:B1_vs_sig_v025}.

\begin{figure}[t!]
\centering
\hspace*{-7mm}
\leavevmode
\epsfysize=5cm 
\epsfbox{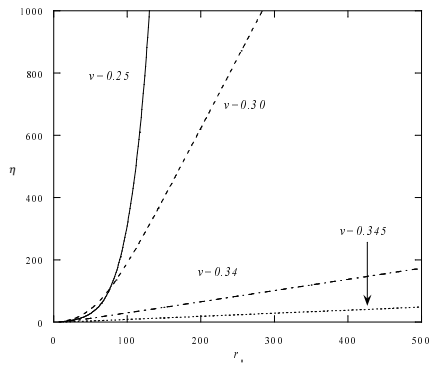}
\\[2mm]
\caption[fig-1]{\label{fig:sig0}
``Zero modes" of the supermassive global monopoles.
We can see that the ``zero modes" are positive and have no zero point.
This indicates the stability of the supermassive
monopoles. At $v=\sqrt{3/8\pi}\approx 0.345$, the supermassive
global monopole coincides with the de Sitter solution, 
which is unstable. This figure shows that the  ``zero modes"
would have zero point as $v \to v_{max}$.
}
\end{figure}

Figure \ref{fig:sig0} shows the ``zero modes" ($\sigma^2=0$), which diverge
and are non-normalizable,
for several values of $v$.
We cannot find
a real zero-mode solution for any value of $B_1$ while there exists a solution
for non-zero positive  $\sigma^2$. 
When a negative eigenmode exists in this kind of linear perturbation analysis,
the perturbed functions have at least one zero point for $\sigma^2=0$
in general.
In our case, ``zero modes" are positive everywhere. This indicates that
the self-gravitating supermassive global monopoles
are stable against the polar perturbations as well as the
spherical perturbations.
As $v$ approaches its maximum value $v_{max}$,
the ``zero modes" become small. It is expected that it
has zero when $v=v_{max}$.
This behavior is consistent with the instability
of the de Sitter solution discussed in Sec.~\ref{sec:ds}.

\section{Discussion}
We investigated the polar stability of the self-gravitating 
supermassive global monopoles and de Sitter solutions in the
Einstein-$O(3)$ scalar system by  the linear perturbation
method \cite{Achucarro}.
Although the de Sitter solutions always have  at least one unstable 
mode for any
value of the VEV of the scalar field, 
the supermassive global monopole solutions do not.
This implies that the supermassive global monopoles
are stable against polar perturbations.
We also find that the minimum eigenvalue is infinitesimal
for the supermassive global monopoles while it becomes non-zero
finite for the ordinary self-gravitating global monopoles and 
non-gravitating counterparts. This is due to the different boundary
conditions at the cosmological horizon.

Our analysis can be   extended to the black hole solution
inside of the monopole, i.e., the black hole solution with
$O(3)$ scalar hair. Such solution was discovered a decade
ago\cite{monopole,tachi} and its supermassive counterpart was recently
discovered by Maison and Liebling\cite{Maison}. It was reported that
these solutions are stable against spherical perturbations.
This result is notable because this scalar hair can be a
physical hair. Hence we should check whether or not this hair is stable
also against polar perturbations.
It should be noted however, that even if this hair is absolutely stable, it 
does not mean the violation of the the black hole no-hair conjecture,
because such conjecture is implicitly assumed in the asymptotic
flatness. Some attempts to extend the conjecture to asymptotically
non-flat spacetime were discussed in Ref.~\cite{toriihair,toriiads}.

\section*{Acknowledgements}

We would like to thank W.Rozycki for correcting the manuscript.



\begin{references}

\bibitem{Barriola}
M. Barriola and A. Vilenkin, Phys. Rev. Lett. {\bf 63}, 341 (1989).

\bibitem{Vilenkin}
A.Vilenkin, Phys. Rev. Lett. {\bf 72}, 3137 (1994).

\bibitem{Linde}
A.D.Linde, Phys. Lett. B {\bf 327}, 208 (1994).

\bibitem{Sakai}
N. Sakai, H. Shinkai, T. Tachizawa and K. Maeda, 
Phys. Rev. D{\bf 53}, 655 (1996).

\bibitem{Liebling}
S. L. Liebling, Phys. Rev. D{\bf 61}, 024030 (1999).

\bibitem{string}
P. Laguna and D. Garfinkle, Phys. Rev. D{\bf 40} 1011 (1989).

\bibitem{MaisonBH}
D. Maison, gr-qc/9912100.

\bibitem{Nucamendi}
U. Nucamendi and D. Sudarsky, Class. Quantum Grav.  {\bf 17}, 4051 (2000).

\bibitem{Derrick} 
G. H. Derrick, J. Math. Phys. {\bf 5}, 1252 (1964).

\bibitem{Goldhaber} 
A. S. Goldhaber, Phys. Rev. Lett. {\bf 63}, 2158 (1989).

\bibitem{Rhie}
S. H. Rhie and D. P. Bennett, Phys. Rev. Lett. {\bf 67}, 1173 (1991).

\bibitem{Perivio}
L. Periviolaropoulos, Nucl. Phys. {\bf B375}, 665 (1992).

\bibitem{Achucarro} 
A. Ach\'ucarro and J. Urrestilla, Phys. Rev. Lett. {\bf 85}, 3091 (2000).

\bibitem{Bennett}
D. P. Bennett and S. H. Rhie, Phys. Rev. Lett. {\bf 65}, 1709 (1990).

\bibitem{Maison}
D. Maison, and S. L. Liebling, Phys. Rev. Lett. {\bf 83}, 5218 (1999).

\bibitem{Breitenlohner}
P. Breitenlohner and D. Z. Freedmann, Phys. Lett. B {\bf 115}, 197 (1982).

\bibitem{torii} 
K. Maeda, T. Tachizawa, T. Torii and T. Maki, Phys. Rev. Lett. {\bf 72}, 450 (1994);
T. Torii, K. Maeda and T. Tachizawa, Phys. Rev. D{\bf 51}, 1510 (1995).

\bibitem{tachi} 
T. Tachizawa, K. Maeda and T. Torii, Phys. Rev. D{\bf 51}, 4054 (1995).

\bibitem{torii2} 
T. Torii, K. Maeda and M. Narita, Phys. Rev. D{\bf 59}, 104002 (1999).

\bibitem{Friedman}
J.L.Friedman, Proc. Roy. Soc. A{\bf 335}, 163 (1973);\\
Chandrasekhar, {\it The Mathematical Theory of Black Holes}
(OXFORD, 1983).

\bibitem{monopole}
K. Lee, V. P. Nair and E. J. Weinberg, Phys. Rev. D{\bf 45}, 2751 (1992);
M. E. Ortiz, Phys. Rev. D{\bf 45}, R2586 (1992);
P. Breitenlohner, F. Forg\'acs, and D. Maison, 
Nucl. Phys. {\bf B383}, 357 (1992);

\bibitem{toriihair}
T. Torii, K. Maeda and M. Narita, Phys. Rev. D{\bf 59}, 064027 (1999).

\bibitem{toriiads}
T. Torii, K. Maeda and M. Narita, Phys. Rev. D{\bf 64}, 044007 (2001).

\end{references}
\end{document}